# Modeling Defects, Shape Evolution, and Programmed Auto-origami in Liquid Crystal Elastomers

Andrew Konya, Vianney Gimenez-Pinto, and Robin Selinger
Liquid Crystal Institute, Kent State University
Kent, OH 44240

Abstract: Liquid crystal elastomers represent a novel class of programmable shape-transforming materials whose shape change trajectory is encoded in the material's nematic director field. Using three-dimensional nonlinear finite element elastodynamics simulation, we model a variety of different actuation geometries and device designs: thin films containing topological defects, patterns that induce formation of folds and twists, and a bas-relief structure. The inclusion of finite bending energy in the simulation model reveals features of actuation trajectory that may be absent when bending energy is neglected. We examine geometries with a director pattern uniform through the film thickness encoding multiple regions of positive Gaussian curvature. Simulations indicate that heating such a system uniformly produces a disordered state with curved regions emerging randomly in both directions due to the film's up/down symmetry. By contrast, applying a thermal gradient by heating the material first on one side breaks up/down symmetry and results in a deterministic trajectory producing a more ordered final shape. We demonstrate that a folding zone design containing cut-out areas accommodates transverse displacements without warping or buckling; and demonstrate that bas-relief and more complex bent/twisted structures can be assembled by combining simple design motifs.

**Introduction**

Liquid crystal elastomers (LCE) undergo reversible shape transformations under any stimulus that changes their degree of nematic order, including heat, illumination, or change of chemical environment (White and Broer, 2015). Complex shape transformation trajectories may be encoded in the material by patterning of the nematic director when the polymer is cross-linked (Liu et al., 2015). The director field is typically controlled by forming samples between flat substrates with a surface anchoring pattern (de Haan et al., 2014). The anchoring pattern can be identical on both substrates, such that for a thin enough sample, the director field is uniform through the material's thickness. Alternatively, the anchoring pattern may differ between the substrates by a prescribed twist angle, where the sense of director twist between the substrates is controlled by the addition of chiral dopant(Sawa et al., 2011). If the material is cross-linked in the nematic phase and the cross-link density is sufficiently high, the director field is "blueprinted" in the material; this programmed pattern then gives rise to locally anisotropic deformation when the scalar nematic order parameter changes, e.g. on heating. As the resulting programmable shape-transformations are reversible, these materials are attractive candidates for applications as conformable soft actuators(Spillmann et al., 2007), to control surface textures(McConney et al., 2013), or to open/close valves or apertures(Modes et al., 2013; Sánchez-Ferrer et al., 2011; Schuhladen et al., 2014).

In previous work, we collaborated with experimenters to explore how even relatively simple director patterns give rise to complex actuation behavior. For instance, a flat, straight LCE ribbon with director twist through the thickness can form either helicoid or spiral structures, depending on sample properties including director twist geometry and sample aspect ratio; remarkably, these chiral structures undergo transitions from right- to left-handed with temperature(Sawa et al., 2011),(Sawa et al., 2013). A larger sample with multiple twisted domains arrayed in stripes gives rise to actuation via accordion folds, while a checkerboard pattern of twisted domains produces a buckled texture(de Haan et al., 2013). In each case, we used finite element simulations to investigate the relationship between director structure and resulting shape actuation. We also examined the soft elastic response of LCE materials with lower cross-link density; these materials show soft elastic response characterized by formation of striped director texture under strain, producing an extended plateau in the stress-strain curve(Mbanga et al., 2010).

Here we explore a variety of programmed shape transformations in LCE's with more complex blueprinted director fields, inspired in many cases by recent experiments. An explanation of simulation methods is provided below in the final section.

**Defects**

Modes and coworkers (Modes et al., 2010) predicted that an initially flat LCE film containing a +1 disclination in the director field, with the director oriented in the azimuthal direction, deforms into the shape of a cone on heating and an anticone (saddle) on cooling. Recent experimental work by Ware, White, McConney and coworkers(McConney et al., 2013),(Ware et al., 2015) demonstrated this behavior and also examined higher order topological defects. Analytical predictions were made in the approximation that bending energy is neglected. We use finite element simulation to examine expected deformation trajectories in the case where bending energy is included.

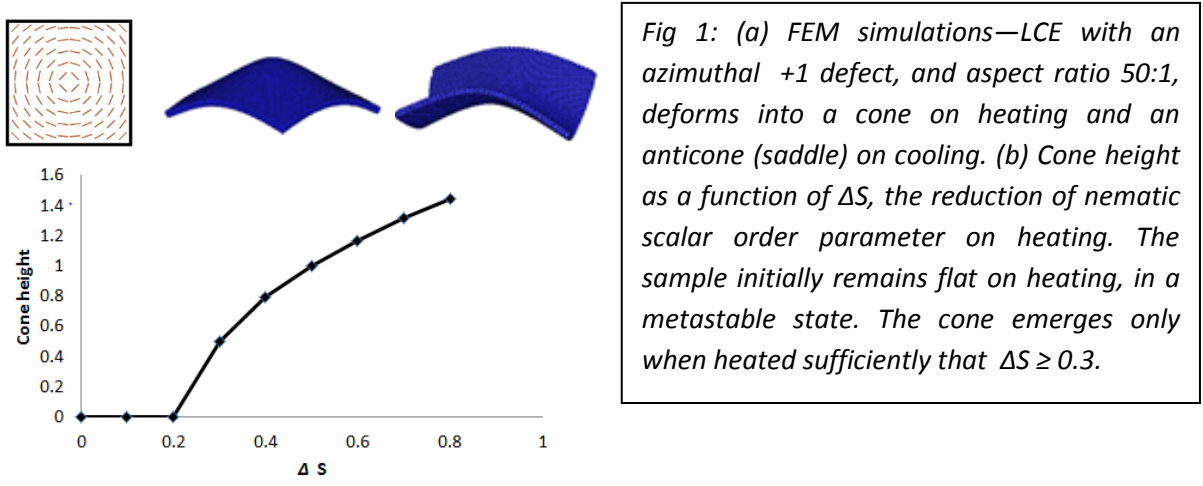

*Fig 1: (a) FEM simulations—LCE with an azimuthal +1 defect, and aspect ratio 50:1, deforms into a cone on heating and an anticone (saddle) on cooling. (b) Cone height as a function of ΔS, the reduction of nematic scalar order parameter on heating. The sample initially remains flat on heating, in a metastable state. The cone emerges only when heated sufficiently that ΔS ≥ 0.3.*

Fig. 1 shows a finite element simulation of an initially square LCE film with an azimuthal +1 defect. The film is represented by a 3-d tetrahedral mesh with an aspect ratio of 50 to 1. As predicted we find formation of a cone on heating, and formation of a saddle on cooling, with respect to the cross-link temperature. We note that on heating, the sample initially remains flat in a metastable state until at a high enough temperature the cone begins to emerge outward, spontaneously breaking up-down symmetry. This behavior is driven by the inclusion of a finite bending energy in the simulation model.

To further investigate effects of finite bending energy, we carry out another simulation of a film containing a +1 defect using a disk-shaped sample; its shape evolution is shown in Fig. 2. The transient structure shows formation of a "sombrero" shape with a distinct rim region, which gradually flattens and snaps through to reach the expected final cone shape. The observed transient shapes may provide insight into structures observed in experiment (de Haan et al., 2012), (McConney et al., 2013).

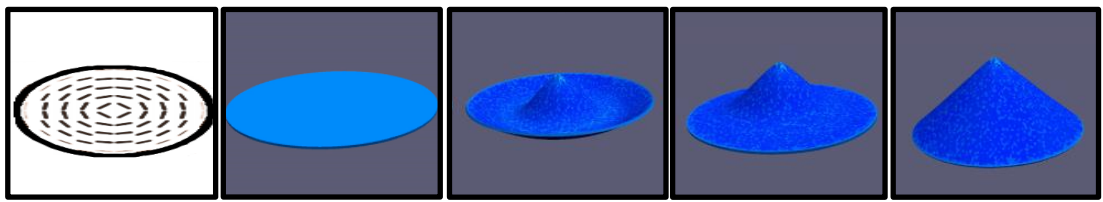

*Fig. 2: Finite element simulation of an LCE disk with a +1 azimuthal defect, on heating. We observe a transient "sombrero" shape with an upward-curving rim. The rim deforms and snaps- through, leaving a cone shaped center surrounded by nearly flat annulus. Eventually the sample reaches the theoretically predicted cone shape* (Modes et al., 2010).

We also investigate shape transformations in LCE films containing higher order defects. Shown in Fig. 3 is a film with a -2 defect. The resulting actuated shape has a protruding center region surrounded by six dimples. Because the top and bottom of the film are symmetric, each of these regions of positive Gaussian curvature has equal probability to pop upwards or downwards. To enhance the probability of finding a symmetric final state, we biased the transformation by applying a temperature gradient at the start of the simulation, heating the sample from one side before stabilizing at uniform temperature.

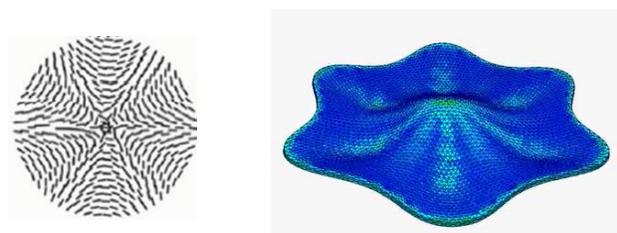

*Fig. 3—Left: structure of a -2 defect. Right: Finite element simulation of an LCE disk containing a -2 defect, shape transformation on heating. Up/down symmetry is broken by heating from one side, producing a symmetric structure. Color indicates local elastic strain energy.*

Fig. 4 (center) shows the final state of an initially flat LCE film containing a -4 defect, again heated from one side to achieve a symmetric structure. Fig 4 (right) shows the structure that forms when the sample is heated uniformly and individual regions pop upwards or downwards at random, producing a disordered and asymmetric shape.

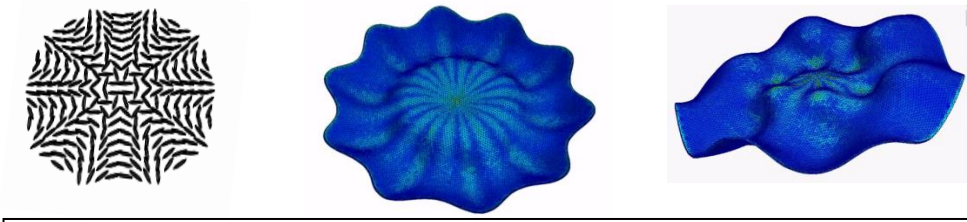

*Fig. 4—Left: Structure of a -4 defect. Center: Finite element simulation of an LCE disk with a -4 defect, shape transformation on heating, heated initially from above. Right: If heated uniformly, each region of positive Gaussian curvature may emerge either upward or downward, producing a complex, disordered final shape.*

In order to achieve deterministic rather than random shape evolution, it is necessary to break this up-down symmetry. Instead of applying a thermal gradient, a better option would be to introduce a material gradient. For instance, the cross-link density may be varied through the thickness of the film for instance by introducing a UV absorbing component in the mixture and cross-link via illumination from one side. Alternatively the director field on top and bottom substrates can be altered so that they are not identical.

**Inducing a fold or twist**

A key goal in design of shape-changing auto-origami devices is the formation of folds, actuated e.g. on heating, induced by a localized gradient in the nematic director field between the top and bottom surfaces. For instance director twist through the thickness of the material can be used to induce a fold(Ware et al., 2015). One challenge is that in addition to driving a folding deformation, an imposed director gradient of this type may also induce a more complex deformation that gives rise to unintended twist or buckling of the sample.

To localize the stresses associated with bending deformation to the hinge region, we designed a structure containing holes. In Fig. 5 we show a bending joint composed of two straight (non-LCE) elastic side walls connected by an array of short LCE beams, each of which has the director field with homeotropic anchoring on top, planar anchoring on bottom parallel to the beam's long axis, with a smooth director gradient in between. On heating, when each beam bends, the accompanying transverse deformation is accommodated by the adjacent voids. As a result, the overall structure bends smoothly without any transverse shrinkage or buckling. This design could function well as an adhesive tape which, when adhered to paper or fabric, induces a fold at a specified temperature. Bending tapes that respond at different temperatures could induce a series of folds that occur in sequence e.g. on gradual warming.

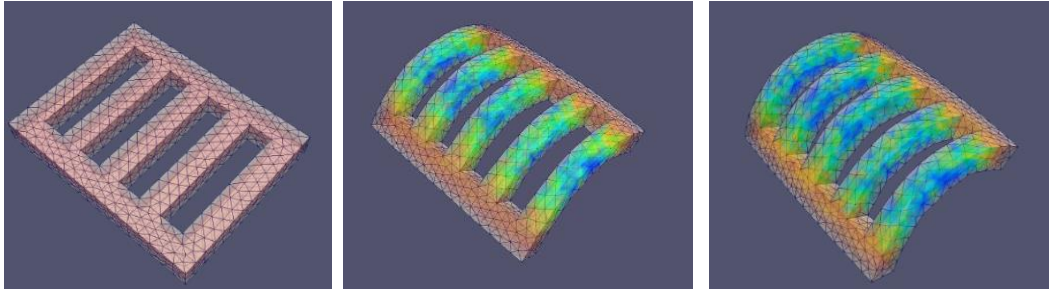

*Fig. 5—Finite Element simulation of an actuator that bends on heating; color indicates strain energy density. Non-LCE side walls are connected by short LCE beams, each of which has homeotropic director orientation along the top surface and planar orientation at the bottom surface with a smooth gradient in between. Open spaces along the bending zone accommodate transverse strain of the LCE beams without lateral shrinkage or buckling of the straight side walls.*

We also demonstrate combining two different types of director gradients along a beam in order to construct a more complex shape evolution. Fig. 6 shows a beam mounted at one end on a substrate, with two different director domains. The left 30% of the sample has a splay geometry with homeotropic anchoring on the bottom and planar anchoring on the top; on heating it curves to form a C-shaped bend. The remaining part of the sample has planar anchoring on both sides with a 90 degree twist between top and bottom surfaces, like those we previously modeled (Sawa et al., 2011),(Sawa et al., 2013), and on heating twists to form a helix. Time evolution on heating produces the trajectory shown in Fig. 6. This structure can be reconfigured reversibly by changing temperature.

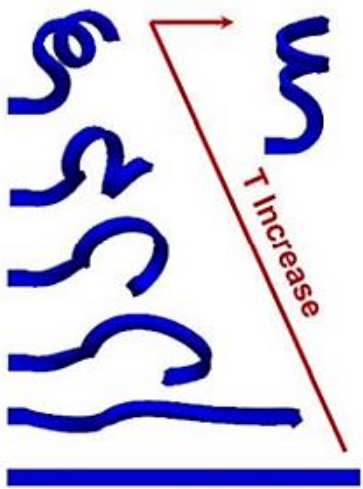

*Fig. 6—Finite Element simulation of shape evolution of an LCE beam on heating. Left 30% of beam has splayed nematic director, which induces a C-shaped bend, while the remaining beam has a 90 degree director twist through the thickness, giving rise to a helical twist.*

**Bas-relief actuation**

Actuators in the form of bas-relief or raised patterns are desirable for applications including haptic displays and for dynamic control of surface texture. A variety of director motifs have been proposed to pattern surfaces to achieve desired surface structures including director fields with (McConney et al., 2013) and without (Mostajeran et al., 2015) topological defects.

Here we examine a simple design motif to drive emergence of a bas-relief pattern on heating. To create a raised region in the shape of a valentine heart, we model a square LCE film with different anchoring conditions on the top and bottom surfaces. On the top surface we impose homeotropic anchoring inside a heart-shaped region, and uniform planar anchoring outside; and on the bottom surface we reverse the two anchoring conditions, as shown in Fig. 6. In between we assume a smooth variation of the nematic director without defects. In practice, it may be necessary to impose a small pre-tilt angle in the homeotropic anchored regions in order to prevent defect formation.

We model shape transformation of this system on heating and find formation of a raised heart-shaped bas-relief structure. Unlike the patterns driven by topological defects of the type discussed above, this design lacks up-down symmetry and thus actuates deterministically with no randomness in the final state.

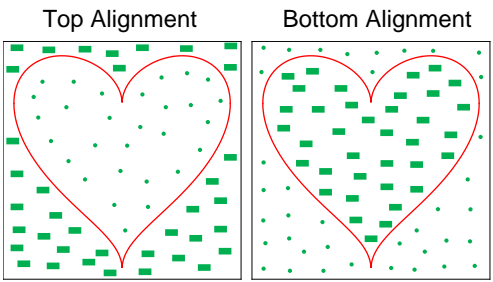

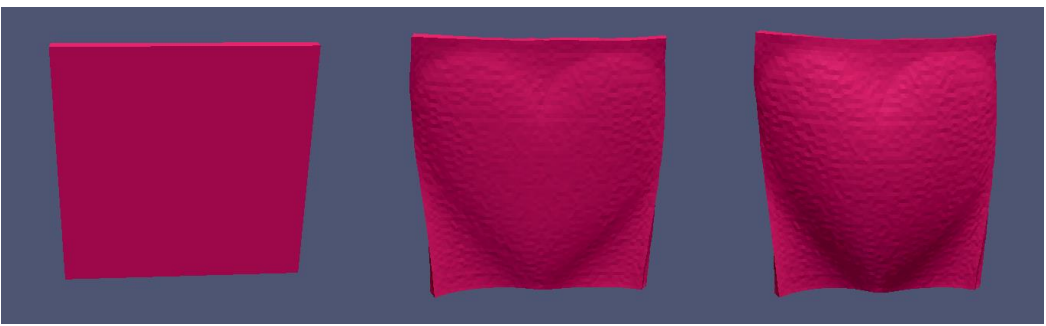

*Fig. 7—Left: Anchoring patterns on top and bottom surfaces with a combination of planar and homeotropic domains. Below: Finite element simulation of actuation on heating of the resulting LCE film; the heart-shaped domain emerges.*

**Simulation Method**

To create a three dimensional non-linear finite element model of our system, we start by meshing our sample in an unstructured tetrahedral mesh using the open-source application Salome (available at http://www.salome-platform.org/.) We write a Hamiltonian for the system with both potential and kinetic energy terms as:

$$U = \sum_{elements\ n}\left[\frac{1}{2}C_{ijkl}\varepsilon_{ij}^n\varepsilon_{kl}^n - \alpha\left(Q_{ij} - Q_{ij}^o\right)_n\varepsilon_{ij}^n\right]V_n + \sum_{nodes\ p}\left[\frac{1}{2}m_p v_p^2\right]$$

Here the first term in the sum over elements is the elastic strain energy, where $\varepsilon_{ij}$ is the Green-Lagrange strain, assumed piece-wise constant over each volume element, and $C_{ijkl}$ is a tensor of elastic constants. The second term in the sum over elements represents coupling between the strain and the nematic order tensor $Q_{ij}$, where $Q_{ij}^o$ represents the "blueprinted" nematic order, defined in each element, at the moment when the sample was cross-linked. The parameter α represents the strength of coupling between nematic order and strain and governs how much the material deforms when the state of nematic order changes, while $V_n$ represents the volume of the element measured in the reference state. The sum over nodes represents calculation of the kinetic energy, where $m_p$ is the effective mass of each vertex, calculated using the lumped mass approximation, and $v_p$ is the node's velocity.

When the sample is heated, we assume that the blueprinted nematic director, described as piecewise constant within each element, remains fixed in the body reference frame but the scalar order parameter drops. Thus the nematic order tensor within each element takes the form $Q_{ij} = S(T)\left[\frac{1}{2}\left(3n_i n_j - \delta_{ij}\right)\right]$, where *S(T)* is the scalar order parameter as a function of temperature and $n_i$ represents the components of the nematic director. Any change in the degree of scalar order thus changes $Q_{ij}$ and drives deformation.

The system evolves forward in time via explicit time stepping in a manner reminiscent of molecular dynamics. At the start of the simulation, all velocities are initially set to zero. At each time step, the effective force on each node in the mesh is calculated as described below and the equations of motion for the nodes are integrated using the velocity Verlet method (Press et al., 2007).

The force calculation proceeds as follows. The displacement field is linear interpolated within each tetrahedral element using the positions of the corner nodes in the current state and in the element's undeformed, stress-free reference state, which we take to be the shape of the element when the material is cross-linked. From the gradients of the displacement field, we calculate the components of the Green-Lagrange strain tensor and from that, calculate the potential energy. In this manner we write the potential energy of the system as a function of the node positions. We calculate the effective force on each of the four nodes by taking the negative derivative of the potential energy within each element with respect to the node positions. The effective force on each node receives a contribution from each element of which it is a member.

Interestingly, our finite element algorithm closely resembles a conventional molecular dynamics simulation (Rapaport, 2004) where nodes in the finite element mesh move instead of atoms. Node interactions take the form of a many body potential, calculated for each element, with energy and forces depending on the node positions both in their current state and in their initial, stress-free reference state. Like a molecular dynamics algorithm, this calculation is entirely local and can thus be easily parallelized for fast execution on a cluster or GPU-equipped processor. If no dissipative forces or mechanical constraints are added, the calculated dynamics conserves the sum of kinetic and potential energy, momentum, and angular momentum. In practice, we add dissipative forces to allow the sample to reach mechanical equilibrium; motion may thus be under- or over-damped. The capability to model momentum-conserving elastodynamics, and not just quasistatic relaxation, allows us to model snap-through shape transitions (Shankar et al., 2013; Smith et al., 2014) and other complex dynamical behavior.

Within this model, if two parts of the sample intersect, they pass through each other, ghostlike, without any collision. To avoid this unphysical behavior in any simulation, we add short-range pairwise repulsive interactions between the surface nodes using the Weeks-Chandler-Andersen potential, which is the Lennard-Jones potential truncated at its minimum and shifted such that the potential vanishes at the cut-off (Ahmed and Sadus, 2009). We then use a molecular dynamics-type algorithm to detect collisions between such surface nodes and calculate the resulting forces.

While less computationally efficient than a two-dimensional shell model (Chung et al., 2015), our fully three-dimensional model can describe material systems with director or cross-link density gradients across the sample thickness. With GPU implementation we can achieve speeds of up to 100 time steps per cpu second for a mesh of more than $10^5$ elements. More details of the model may be found in (Mbanga, 2012) and (Gimenez-Pinto, 2014).

**Discussion**

Devices engineered using patterned LCEs have a somewhat limited design space as the nematic director field may be controlled only via anchoring patterns at the sample's surface. For this reason, an arbitrary three-dimensional director field cannot be realized. An experimental method to blueprint LCE materials with fully three-dimensional voxel-by-voxel control of the director orientation would provide a broader class of available structures, but has not yet been developed to the best of our knowledge.

By contrast, other materials with anisotropic shape evolution have recently been introduced (Gladman et al., 2016) that can be printed in successive layers, providing a greater degree of control over the full three-dimensional orientation field. Finite element modeling will play a key role in engineering both types of devices. Future developments will include numerical methods to address the so-called inverse problem, that is, to design a director field that gives rise to a shape actuation that closely follows a prescribed trajectory and achieves a target final shape.

**Conclusion**

Finite element simulation is a valuable design tool to explore a wide variety of device designs for materials undergoing programmable shape change. The three-dimensional nature of our model allows study of systems where bending energy plays an important role, and allows modeling of samples where the nematic director is not uniform through the film thickness.

While quasistatic relaxation is sufficient to model a slow shape evolution process, some shape transitions involve mechanical instabilities and may show rapid snap-through behavior. These rapid shape transitions can be modeled via underdamped, momentum-conserving elastodynamics simulation. We plan to further explore such systems in future work.

**Acknowledgements**: The authors thank Tim White, Mike McConney, Qihuo Wei, and Badel Mbanga for helpful discussions. Work supported by the National Science Foundation NSF-DMR 1106014, NSF-DMR-1409658, and NSF-CMMI 1436565. Computer resources provided by the Ohio Supercomputer Center and by the Kent State College of Arts and Sciences.